\documentclass[aps,preprint,nofootinbib]{revtex4}%
\usepackage{amsfonts}
\usepackage{amsmath}
\usepackage{amssymb}
\usepackage{lipsum}
\usepackage{xcolor}
\usepackage{amssymb}
\usepackage{amsmath}
\usepackage{slashed}%notacion Feynmann
\usepackage{float}
\usepackage{physics}
\usepackage{graphicx}
\usepackage{hyperref}
\hypersetup{colorlinks}
\usepackage{cancel}%paquete para cancelar expresiones
\usepackage[utf8]{inputenc}
\usepackage{stackrel}%paquete para poner símbolos arriba de los caracteres 
\usepackage{caption}

\newcommand\blfootnote[1]{%
  \begingroup
  \renewcommand\thefootnote{}\footnote{#1}%
  \addtocounter{footnote}{-1}%
  \endgroup
}
\usepackage{float}
\usepackage{graphicx}%
\providecommand{\U}[1]{\protect\rule{.1in}{.1in}}
\begin{document}
%%%%%%%
 
\title{Quasitopological electromagnetism: Reissner-Nordstr\"om black strings in Einstein and Lovelock gravities}

\author{$^{1,2 , \star}$Adolfo Cisterna, $^{3,\dagger}$Carla Henr\'iquez-B\'aez, $^{3,\ddagger}$Nicol\'as Mora, $^{3,*}$ Leonardo Sanhueza}

\affiliation{$^{1}$Dipartimento di Fisica, Universita di Trento, Via Sommarive 14, 38123 Povo, Trentino, Italy.}

\affiliation{$^{2}$TIFPA - INFN, Via Sommarive 14, 38123 Povo, Trentino, Italy,}

\affiliation{$^{3}$Departamento de F\'isica, Universidad de Concepci\'on, Casilla 160-C, Concepci\'on, Chile}

\blfootnote{$\star$\url{adolfo.cisterna@unitn.it}\\$^{\dagger}$\url{carlalhenriquez@udec.cl}\\$\ddagger$\url{nicolasmora@udec.cl}\\$^{*}$\url{lsanhueza@udec.cl}
}

\begin{abstract}

In this work, we provide consistent compactifications of Einstein-Maxwell and Einstein-Maxwell-Lovelock theories on direct product spacetimes of the form $\mathcal{M}_D=\mathcal{M}_d\times\mathcal{K}^{p}$, where $\mathcal{K}^p$  is a Euclidean internal manifold of constant curvature. For these compactifications to take place, the distribution of a precise flux of  $p$-forms over the internal manifold is required. The dynamics of the $p$-forms are demanded to be controlled by two types of interactions: first, by specific couplings with the curvature tensor and, second, by a suitable interaction with the electromagnetic field of the $d$-dimensional brane, the latter being dictated by a modification of the recently proposed theory of quasitopological electromagnetism. The field equations of the corresponding compactified theories, which are of second order, are solved and general homogenously charged black $p$-branes are constructed. We explicitly provide homogenous Reissner-Nordstr\"om  black strings and black $p$-branes in Einstein-Maxwell theory and homogenously charged Boulware-Deser black $p$-branes for quadratic and cubic Maxwell-Lovelock gravities.

\end{abstract} 

\maketitle

\section{Introduction}

From a black hole in $d$ dimensions, in some cases, it is possible to construct a black hole in a higher dimension $D=d+p$ by simply adding $p$ extra flat directions onto the original spacetime.
Such solutions are known as black strings or black $p$-branes. Although they seem to be trivial, there is a non-negligible amount of physics underlying them \cite{Obers:2008pj,Harmark:2007md}. 
The canonical example is the Schwarzschild black hole, which can be trivially extended to a black brane in any dimension $D=4+p$. 
Schwarzschild black branes are connected with more sophisticated spacetimes since they emerge as the large angular momentum limits of the five-dimensional black ring \cite{Emparan:2001wn} and the six-dimensional Myers-Perry black hole \cite{Myers:1986un}. In addition, they are thermodynamically disfavored with respect to the Tangherlini black hole \cite{Tangherlini:1963bw} and in consequence were shown to suffer from the so-called Gregory-Laflamme (GL) instability \cite{Gregory:1993vy}\footnote{The GL instability goes beyond the domain of general relativity, persisting for example, in theories with higher curvature corrections \cite{Giacomini:2015dwa,Giacomini:2016ftc,Lagos:2017vap}.}. Numerical studies suggest that for $d<13$ the final state of this instability is given by a chain of black holes and naked singularities that interpolates along the extended direction \cite{Lehner:2010pn}, exposing in this manner the permeability of the cosmic censorship conjecture in higher dimensions \cite{Hawking:1973uf}.\\
Whether or not homogenous black $p$-branes belong to the black hole spectrum of a given theory depends on the ability of the theory to support compactifications over direct product spacetimes of the form $\mathcal{M}_D=\mathcal{M}_d\times\mathcal{K}^p$, with $\mathcal{M}_d$ a spacetime of dimension $d$ and $\mathcal{K}^p$ a $p$-dimensional internal manifold of constant curvature. In the Kaluza-Klein nomenclature \cite{Kaluza,Klein} this means a compactification in which the vector field vanishes and the dilaton field is a constant \cite{Overduin:1998pn,Duff:1986hr}. Despite the inherent simplicity of these compactifications in general relativity (GR), the mere inclusion of one extra scale into the theory, a cosmological constant, produces unbearable incompatibilities between the field equations on the brane and on the internal manifold. As a consequence, no homogenous Schwarzschild (anti)-de Sitter [(A)dS] black strings emerge\footnote{One way to circumvent these obstructions is to consider warped spacetimes \cite{Chamblin:1999by}. These solutions are known as inhomogenous black strings, and they do exist in Lovelock theories as well \cite{Kastor:2006vw,Giribet:2006ec,Kastor:2017knv}. Similar constructions can be found in \cite{Kanti:2018ozd,Nakas:2019rod,Nakas:2020crd}.}.

 Considering internal manifolds with a nonzero constant curvature slightly modify the picture, nevertheless, it is not possible to accommodate a small cosmological constant while keeping small extra dimensions. 
To circumvent these difficulties, a well-known approach is to consider higher-dimensional matter fields, in concrete minimally coupled $(p-1)$-forms fields $A_{[p-1]}$ that dress the corresponding internal manifold through a field strength $F_{[p]} = dA_{[p-1]}$ which is proportional to its volume form \cite{Freund:1980xh,RandjbarDaemi:1982hi}. 
For compactifications over flat internal manifolds, $0$-forms were proven to effectively support black strings solutions providing at the same time a simple setting in which to analyze their mechanical instability \cite{Cisterna:2017qrb}\footnote{These black strings were shown to be mechanically stable under generic perturbations of the metric and the scalar field. For such perturbations, the scalar field dressing forbids the presence of the GL mode \cite{Cisterna:2019scr}.}$^{,}$\footnote{In the $3+1$-dimensional case nonlinear sigma models have been particulary useful in the construction of BTZ black strings \cite{Astorino:2018dtr, Giacomini:2019qov}.}. 
These black strings were also constructed for generic Lovelock theories, providing the black string version of the Boulware-Deser black hole \cite{Cisterna:2018mww}.\\
For the theory to be compatible notwithstanding the existence of the higher curvature terms, the scalar field dynamic must be controlled by proper nonminimally coupled interactions in which the scalars are kinetically coupled with Lovelock tensors\footnote{Those interactions naturally emerge in Horndeski theory \cite{Horndeski:1974wa}. Other higher curvature theories has been explored along these lines \cite{Cisterna:2018jsx,Arratia:2020hoy}.}. This approach paved the way to pursue the compactification of Lovelock theory on internal manifolds with a nontrivial curvature \cite{Cisterna:2020kde}. 
In order to do so, and in analogy with the scalar case, $p$-form fields must be coupled with curvature tensors. Such a setting is naturally provided in the vector-tensor theory proposed in \cite{Horndeski:1976gi}, which provides the most general second-order vector-tensor field equations compatible with the principle of charge conservation and that when going to a Minkowski background reproduce Maxwell equations.
This model, whose more general form has been found in \cite{Feng:2015sbw}, allows the compactification of generic Lovelock theories. Moreover, it not only allows for the construction of Lovelock black strings but also the construction of Schwarzschild black strings, this by providing a natural arena in which a generic Lovelock theory is compactifiable into Einstein gravity \cite{Canfora:2021ttl}. \\
The inclusion of matter fields on the brane makes compactifications much more unlikely, in fact, for example, no simple Reissner-Nordstr\"om black string is known. 
In this respect, charged black $p$-branes have been previously explored in the context of string theory \cite{Horowitz:1991cd}, for a specific Einstein-Maxwell-dilaton theory appearing from the low energy limits of ten-dimensional string theories. These solutions are homogeneous. Nevertheless, the dilaton and gauge fields are coupled and there is no limit that brings the black hole on the brane into the Reissner-Nordstr\"om solution. 
On the other hand, it has been shown that pure Lovelock theories of order $k$ support homogenous black strings constructed from charged black holes as long as the brane electromagnetic field 
is given by a $p$-form field for which $p$ equals $k$ \cite{Giacomini:2018sho}. Then, electrically charged black strings are present only for pure Gauss-Bonnet theory, with no Reissner-Nordstr\"om limit, while for higher-order Lovelock models only magnetically charged solutions might arise, depending on the dimension of the black hole horizon located on the brane.\\
In this work, we propose to investigate the compactification of Einstein-Maxwell and Einstein-Maxwell-Lovelock theories on direct product spacetimes of the form discussed above and in this manner to provide homogenously charged black strings, specifically, the black $p$-brane extensions of the Reissner-Nordstr\"om and charged Boulware-Deser black holes. 
We start with Einstein-Maxwell theory, which can be properly compactified by considering the direct coupling of the brane electric field with magnetic $p$-forms living on the internal manifold, this along the lines of a submodel of \cite{Feng:2015sbw} recently dubbed quasitopological electromagentism \cite{Liu:2019rib,Cisterna:2020rkc}\footnote{These theories should not to be confused with the recently proposed models of electromagnetic quasitopological gravities \cite{Cano:2020qhy,Bueno:2021krl}.}.
Once the compactification is achieved, we move into the construction of the corresponding black $p$-branes solutions. In analogy with this case, and following the procedure developed in \cite{Cisterna:2020kde}, we move into decompactification of Einstein-Maxwell-Gauss-Bonnet and Einstein-Maxwell-cubic Lovelock theories, providing homogenously charged Boulware-Deser black $p$-branes.\\
The paper is organized as follows: Section II is devoted to presenting the model and the corresponding field equations.
In Sec. III we proceed with the compactification of Einstein-Maxwell theory on a $(d+p)$-dimensional direct product spacetime. We show how the compatibility between the field equations is reached and how the construction of black $p$-branes with one or multiple internal manifolds is achieved. 
Section IV is destined to apply the techniques of Sec. III into Einstein-Maxwell-Gauss-Bonnet and Einstein-Maxwell-cubic Lovelock theories constructing the corresponding charged Boulware-Deser extended solutions. We finally outline some conclusions in Sec. V.

\section{Model and field equations}

Lovelock theory \cite{Lovelock:1971yv} corresponds to the natural generalization of GR in higher dimensions. Its Lagrangian of order $k$ is defined by
\begin{equation}
\mathcal{L}_{Lovelock}=\sum_{k=0}^{[D/2]}\alpha_{k}\mathcal{L}^{(k)}, 
\end{equation}
where the $\alpha_k$ are the Lovelock couplings and $\mathcal{L}^{(k)}$ the topological Euler densities defined by
\begin{align}
\mathcal{L}^{(k)}=\sum^{[D/2]}_{k=0}\frac{1}{2^{k}}\delta^{A_1\cdots A_{2k}}_{B_1\cdots B_{2k}}R^{B_1B_2}{}{}_{A_{1}A_{2}}\cdots R^{B_{2k-1}B_{2k}}{}{}_{A_{2k-1}A_{2k}}.
\end{align}
Here, $\delta_{B_1...B_{2k}}^{A_1...A_{2k}}$ stands for the antisymmetric generalized Kronecker delta. In addition, each Euler density contributes nontrivially to the field equations for $k<[D/2]$, otherwise, they reduce to a boundary term or identically vanish. Einstein gravity plus a cosmological constant is obtained by truncating the series up to first order. In the next sections, we use Lovelock Lagrangians up to the cubic order, which are  
\begin{align}
\alpha_0\mathcal{L}^{(0)}&=-2\Lambda\\
\alpha_1\mathcal{L}^{(1)}&=R\\
\alpha_2\mathcal{L}^{(2)}&=\alpha_2(R^2-4R_{AB}R^{AB}+R_{ABCD}R^{ABCD})\\
\alpha_3\mathcal{L}^{(3)}&=\alpha_3(R^{3}-12 R R_{AB}R^{AB}+16 R_{AB}R^{A}{}_{C}R^{BC}+24R_{AC}R_{BD}%
R^{ABCD}+3R R_{ABCD}R^{ABCD}\nonumber\\
&  -24R_{AB}R^{A}{}_{CDE}R^{BCDE}+4R_{ABCD}R^{ABEF}R^{CD}{}_{EF}-8R_{ABC}%
{}^{D} R^{AECF}R^{B}{}_{EDF}\ ), 
\end{align}
where we have defined $\alpha_0=-2\Lambda$ and we have set $\alpha_1=1/16\pi G=1$.
In order to perform compactifications of Einstein-Maxwell theory or any Lovelock-Maxwell theory on spaces of the form $\mathcal{M}_D=\mathcal{M}_d\times\mathcal{K}^p$, first, we  need to get rid of the incompatibilities that naturally emerge from the field equations once they are projected on the brane and the corresponding internal manifold. 
These incompatibilities naturally emerge due to the presence of both the Maxwell term and the higher curvature contributions that Lovelock gravity brings into the theory. 
Aiming to circumvent these obstructions and in order to perform the compactifications, we cure these incompatibilities by adding extra matter fields, in particular, specific higher-dimensional $p$-form fields whose dynamics naturally contribute to the Maxwell energy-momentum tensor and to the geometric tensors coming from the Lovelock Lagrangian. 
For this to be the case, these $p$-forms are required to be proportional to the volume form of the corresponding internal manifold $\mathcal{K}^p$ providing a suitable dressing that might solve the incompatibilities. 
The dynamic of these matter fields must be given by a model that couples curvature tensors and $p$-form fields in such a manner that the corresponding field equations are of second order. These features are properly englobed by the theory presented in \cite{Feng:2015sbw} where, in complete analogy with Lovelock gravity, the Lagrangian is built in terms of polynomial invariants constructed from curvature tensors and $p$-form fields. Its Lagrangian is given by 
\begin{align}
\mathcal{L}^{(k,n)}_{p-forms}\left[g, B_{[p-1]}\right]=\sum^{[D/2]}_{k=0}\sum^{[D/p]}_{n=0}\frac{\beta_{k}}{2^k(p!)^{n}}&\delta^{A_1\cdots A_{2k}C_1^{1}\cdots C_1^{p}\cdots C_n^{1}\cdots C_n^{p}}_{B_1\cdots B_{2k}D_1^{1}\cdots D_1^{p}\cdots D_n^{1}\cdots D_n^{p}}R^{B_1B_2}{}{}_{A_1A_2}\cdots R^{B_{2k-1}B_{2k}}{}{}_{A_{2k-1}A_{2k}}\nonumber\\
&\times Z^{D_1^{1}\cdots D_1^{p}}{}_{C_1^{1}\cdots C_1^{p}}\cdots Z^{D_n^{1}\cdots D_n^{p}}{}_{C_n^{1}\cdots C_n^{p}} \label{eq:general-action-matter},
\end{align}
where $Z$ represents the following bilinear combination:
\begin{align}
Z^{A_1\cdots A_p}{}_{B_1\cdots B_p}&:=M^{A_1\cdots A_p}M_{B_1\cdots B_p},\label{eq:Z-tensor}
\end{align}
for, up to this point, an arbitrary electromagnetic field strength $M_{[p]}=dB_{[p-1]}$. Here, $M$ or, consequently, $B$ might represent, depending on the case under consideration,  either electric or magnetic configurations while $\beta_k$ represents dimensionful coupling constants. For electric configurations we denote $M$ as $F_{AB}$ while for magnetic configurations it will be denoted as $H_{A_1...A_p}$.
When $M$ corresponds to the 2-form electromagnetic field strength $F$ and only one curvature tensor is considered, this Lagrangian  precisely reproduces the model proposed in \cite{Horndeski:1976gi}.
\\
On the other hand, the second-order character of the field equations is ensured by the Bianchi identity that $M_{A_1...A_p}$ satisfies along with a similar relation imposed on the $Z$ tensor (for details see \cite{Feng:2015sbw}).
It is interesting to remark that, contrary to the case of Lovelock theories, these densities do not represent topological invariants in the critical dimension. Conversely they do contribute to field equations when $k\geq\left[D/2\right]$.\\ 
The field equations of Einstein-Lovelock-Maxwell theory supplemented with the interaction defined by (\ref{eq:general-action-matter}) take the form
\begin{align}
\sum^{[D/2]}_{k=0}\left[\alpha_{k}E^{\left(k\right)}_{AB}-\beta_{k}T^{\left(k-1, 1\right)}_{AB,p}\right]&=T^{em}_{AB},\label{eq:field-equations-action-theory}
\end{align}
where $E^{\left(k\right)}_{AB}$ is the Lovelock tensor of order $k$, 
\begin{align}
E^{\left(k\right)}_{AB}&=-\frac{1}{2^{k+1}}g_{\left(A\right|C}\delta^{CA_{1}\cdots A_{2k}}_{\left|B\right)B_1\cdots B_{2k}}R^{B_1B_2}{}{}_{A_{1}A_{2}}\cdots R^{B_{2k-1}B_{2k}}{}{}_{A_{2k-1}A_{2k}}\label{eq:lovelock-tensor-p},
\end{align}
and $T^{\left(k, 1\right)}_{AB,p}$ is the corresponding energy-momentum tensor associated to (\ref{eq:general-action-matter}), namely, 
\begin{align}
T^{\left(k, 1\right)}_{AB,p}=&\frac{1}{2^{k+1}p!}g_{AB}\delta^{A_1\cdots A_{2k}C_1\cdots C_p}_{B_1\cdots B_{2k}D_1\cdots D_p}R^{B_1B_2}{}{}_{A_{1}A_{2}}\cdots R^{B_{2k-1}B_{2k}}{}{}_{A_{2k-1}A_{2k}}Z_{}^{D_1\cdots D_p}{}_{C_1\cdots C_p}\nonumber\\
&-\frac{k}{2^{k}p!}\delta^{A_1A_2\cdots A_{2k}C_1\cdots C_p}_{B_1\left(A\right|\cdots B_{2k}D_1\cdots D_p}R^{B_1}{}_{\left|B\right)A_1A_2}R^{B_3B_4}{}{}_{A_3A_4}\cdots R^{B_{2k-1}B_{2k}}{}{}_{A_{2k-1}A_{2k}}Z_{}^{D_1\cdots D_p}{}_{C_1\cdots C_p}\nonumber\\
&+\frac{2k}{2^{k}p!}\delta^{A_1\cdots A_{2k}C_1\cdots C_p}_{\left(A\right|\cdots B_{2k}D_1\cdots D_p}g_{A_2\left|B\right)}R^{B_3B_4}{}{}_{A_3A_4}\cdots R^{B_{2k-1}B_{2k}}{}{}_{A_{2k-1}A_{2k}}\nabla_{A_1}M_{}^{D_1\cdots D_p}\nabla^{B_2}M^{}_{C_1\cdots C_p}\nonumber\\
&+\frac{2pk}{2^{k}p!}\delta^{A_1\cdots A_{2k}C_1\cdots C_p}_{\left(A\right|\cdots B_{2k}D_1\cdots D_p}g_{A_2\left|B\right)}R^{B_3B_4}{}{}_{A_3A_4}\cdots R^{B_{2k-1}B_{2k}}{}{}_{A_{2k-1}A_{2k}}R^{D_1}{}_{E}{}^{B_2}{}_{A_1}Z_{}^{ED_2\cdots D_p}{}_{C_1\cdots C_p}\nonumber\\
&-\frac{p}{2^{k}p!}\delta^{A_1\cdots A_{2k}C_1\cdots C_p}_{B_1\cdots B_{2k}\left(A\right|\cdots D_p}R^{B_1B_2}{}{}_{A_{1}A_{2}}\cdots R^{B_{2k-1}B_{2k}}{}{}_{A_{2k-1}A_{2k}}Z_{\left|B\right)}{}^{D_2\cdots D_p}{}_{C_1\cdots C_p},\label{eq:energy-momentum-tensor-k-p}
\end{align}
and $T^{em}_{AB}$ is the standard Maxwell-Faraday tensor
\begin{equation}
T^{em}_{AB}=\frac{1}{2}(F_{AC}F_B^C-\frac{1}{4}g_{AB}F_{CD}F^{CD}).
\end{equation}
Variations with respect to the gauge fields $A_{[p-1]}$ and $B_{[p-1]}$, respectively, provide
\begin{align}
\nabla_AF^{AB}&=0,\\
\sum^{\left[D/2\right]}_{k=0}\frac{\beta_{k}}{2^{k-1}}\delta^{A_1\cdots A_{2k}C_1\cdots C_p}_{B_1\cdots B_{2k}D_1\cdots D_p}R^{B_1B_2}{}{}_{A_1A_2}\cdots R^{B_{2k-1}B_{2k}}{}{}_{A_{2k-1}A_{2k}}\nabla^{D_1}M_{C_1\cdots C_p} &= 0.
\end{align}
Note that here we have considered the case in which 
$n=1$, i.e. the gauge field equation for $M$ is free of nonlinearities\footnote{Higher values of $n$ might be useful when compactifying nonlinear electrodynamics of the form $\left(-\frac{1}{4}F_{AB}F^{AB}\right)^q$, with $q$ some natural.}. \\
In the next sections, we address the compactification of Einstein-Maxwell and Einstein-Lovelock-Maxwell theories, this by supplementing our theories with suitable models contained in (\ref{eq:general-action-matter}). As already mentioned, compactification in the absence of the Maxwell field has been systematically performed in \cite{Cisterna:2020kde}. However, under the presence of the Maxwell action, this approach needs to be slightly modified. The Maxwell action does not introduce incompatibilities regarding the presence of the curvature terms, but it does due to the presence of terms of the form $F_{AB}F^{AB}$. Then, it is mandatory to consider (\ref{eq:general-action-matter}) in the case in which $k=0$. Moreover, to properly get rid of the incompatibilities generated by the Maxwell term, the $Z$ bilinear tensor is forced to be inhomogeneous,
\begin{equation}
Z^{A_1A_2}{}_{B_1...B_p}=F^{A_1A_2}H_{B_1...B_p}, \label{IZ}
\end{equation}
being composed by the original electric Maxwell field strength, which lives on the brane $\mathcal{M}_d$ and the magnetic $p$-form field strength $H_{B_1...B_p}$, which is demanded to live exclusively on the internal manifold $\mathcal{K}^p$. This theory, which is contained in (\ref{eq:general-action-matter}), has been dubbed quasitopological electromagnetism (QTE) and has been recently explored in the context of black hole solutions and cosmology \cite{Liu:2019rib,Cisterna:2020rkc}.\\
In the remaining sections of this work, we work in the compactification of Einstein-Lovelock-Maxwell theories on direct product spaces with geometries better represented by 
\begin{equation}
d s^{2}=g_{A B} d x^{A} d x^{B}=\tilde{g}_{\mu \nu}(y) d y^{\mu} d y^{\nu}+\hat{g}_{i j}(z) d z^{i} d z^{j},  \label{metric}
\end{equation}
where $\tilde{g}_{\mu \nu} d y^{\mu} d y^{\nu}$ stands for the $d$-dimensional spacetime manifold $\mathcal{M}_{d}$ while $\hat{g}_{i j}(z) d z^{i} d z^{j}$
represents a $p$ -dimensional Euclidean manifold $\mathcal{K}^{p}$,
\begin{equation}
\hat{g}_{i j}(z) d z^{i} d z^{j}=\frac{d \vec{z} \cdot d \vec{z}}{\left(1+\frac{\gamma}{4} \sum_{j=1}^{p} z_{j}^{2}\right)^{2}}
\end{equation}
of constant curvature,
\begin{equation}
\hat{R}_{i j k l}=\gamma\left(\hat{g}_{i k} \hat{g}_{j l}-\hat{g}_{i l} \hat{g}_{j k}\right), \label{cuv}
\end{equation}
with $\gamma$ defining the corresponding curvature radius $R_0=|\gamma|^{-1}$. Tildes denote quantities living on the brane $\mathcal{M}_d$, while hats correspond to quantities living over the internal manifold $\mathcal{K}^p$. In addition, and in accordance with this geometry the configurations for $F$ and $H$ are, respectively, given by
\begin{align}
\tilde{F}_{\mu\nu}&=-a^{\prime}(r)\delta_{\mu\nu}^{tr}\\
\hat{H}_{i_{1} \cdots i_{p}}&=\frac{p!Q_{m}}{\left(1+\frac{\gamma}{4} \sum_{j=1}^{p} z_{j}^{2}\right)^{p}} \hat{\delta}_{i_{1} \cdots i_{p}}^{z_1\cdots z_2}, \label{fieldans}
\end{align}
with $Q_m$ regarded as a generalized magnetic charge. 

\section{Compactification of Einstein-Maxwell theory}

We start focusing on the Einstein-Maxwell case. In this case no extra higher curvature terms appear and, in consequence, no incompatibilities associated to those types of interactions. Thus, to make the theory compatible under the compactification, we complement Einstein-Maxwell theory with (\ref{eq:general-action-matter}) for the case in which $k=0$ and $Z$ given by (\ref{IZ}). The resulting interaction, which we dubbed  the QTE Lagrangian, is given by 
 \begin{equation}
\mathcal{L}_{QTE}=-2p!\beta_0 F_{A_1A_2}F^{A_1A_2}H_{B_1...B_p}H^{B_1...B_p}.  \label{QTE}
\end{equation}
%The first  observation to be made is that, for a purely electric field $F_{AB}$ and a purely magnetic field $H_{A_1...A_p}$ living on the internal manifold, Lagrangian (\ref{QTE}) possesses only one non-trivial term once it is expanded, i.e
%\begin{equation}
%-\frac{1}{2p!}H_{A_1\dots A_p}H^{A_1\dots A_p}
%\mathcal{L}_{QTE}=-\frac{1}{2p!}H_{A_1\dots A_p}H^{A_1\dots A_p}-2p!\beta_0 F_{A_1A_2}F^{A_1A_2}H_{B_1...B_p}H^{B_1...B_p} \label{lag}, 
%\end{equation}
This yields an action principle of the form 
\small
\begin{equation}
S=\int{(R-2\Lambda-\frac{1}{4}F_{AB}F^{AB}-\frac{1}{2p!}H_{A_1\dots A_p}H^{A_1\dots A_p}-2p!\beta_0 F_{A_1A_2}F^{A_1A_2}H_{B_1...B_p}H^{B_1...B_p})\sqrt{-g}d^{d+p}x},
\end{equation}
\normalsize
where, in order to properly account for the presence of a nontrivial cosmological constant, we have also included a kinetic term for $H_{[p]}$. 
Variation with respect to the metric delivers Einstein field equations 
\begin{equation}
G_{AB}+g_{AB}\Lambda-T^{em}_{AB}-T^{H}_{AB}-\beta_0T^{QTE}_{AB}=0,
\end{equation} 
where
\begin{eqnarray}
T^{H}_{AB}&=&\frac{1}{2\left(p-1\right)!}H_{A A_{2}\ldots A_{p}}H_{B}{}^{A_{2}\ldots A_{p}}-\frac{1}{4p!}g_{AB}H^2 \ , \nonumber \\
T^{QTE}_{AB}&=&2p!(2F_{AC}F_B^CH^2+pH_{AA_2...A_p}H_B^{A_2...A_p}F^2-\frac{1}{2}g_{AB}F^2H^2), \label{TMNU}
\end{eqnarray}
while variations with respect to the electromagnetic potentials provide the following gauge field equations  
\begin{eqnarray}
\nabla_A[(1+8p!\beta_0H^2)F^{AB}]&=&0, \label{gaugeq6}\\
\nabla_{A_{1}}[(1+4p!^2\beta_0F^2)H^{A_1...A_p}]&=&0 \label{gaugeq7}.
\end{eqnarray}
It is direct to observe that, for a magnetic $p$-form with a field strength proportional to the volume for of the internal space (\ref{fieldans}), the associated gauge field equation (\ref{gaugeq7}) is automatically satisfied, while the gauge field equation associated with the electric field (\ref{gaugeq6}) is satisfied provided that
\begin{equation}
a(r)=-\frac{1}{(d-3)}\frac{Q_e}{r^{(d-3)}},
\end{equation}
with $Q_e$ and the integration constant related with the electric charge. 
Due to the nature of the product spacetime $\mathcal{M}_D$, ultimately expressed in the form of the metric ansatz (\ref{metric}), the Einstein field equations are directly projectable on $\mathcal{M}_d$ and $\mathcal{K}^p$, thus,  
\footnotesize
\begin{align}
&\tilde{R}_{\mu\nu}-\frac{1}{2}\tilde{g}_{\mu\nu}(\tilde{R}_d+\hat{R}_p)+\tilde{g}_{\mu\nu}\Lambda-\frac{1}{2}\left(\tilde{F}_{\mu\lambda}\tilde{F}_{\nu}{}^{\lambda}-\frac{1}{4}\tilde{g}_{\mu\nu}\tilde{F}^2\right)+\frac{1}{4p!}\tilde{g}_{\mu\nu}\hat{H}^2-2p!\beta_0\left(2\tilde{F}_{\mu\lambda}\tilde{F}_{\nu}{}^{\lambda}-\frac{1}{2}\tilde{g}_{\mu\nu}\tilde{F}^2\right)\hat{H}^2=0,\nonumber\\ 
&\hat{R}_{ij}-\frac{1}{2}\hat{g}_{ij}(\tilde{R}_d+\hat{R}_p)+\hat{g}_{ij}\Lambda+\frac{\tilde{F}^2}{8}+\left(\frac{\hat{H}_{ii_2...i_p}\hat{H}_{j}{}^{i_2...i_p}}{2(p-1)!}-\frac{1}{4p!}\hat{g}_{ij}\hat{H}^2\right)-2p!\beta_0\left(p\hat{H}_{ii_2...i_p}\hat{H}_{j}{}^{i_2...i_p}-\frac{1}{2}\hat{g}_{ij}\hat{H}^2\right)\tilde{F}^2=0. \label{EOMQTE}
\end{align}
\normalsize
Here, we make use of the shorthand notations $\tilde{F}^2=\tilde{F}_{\mu\nu}\tilde{F}^{\mu\nu}$ and $\hat{H}^2=\hat{H}_{i_1..i_p}\hat{H}^{i_1..i_p}$.
In addition, the Ricci scalar of the whole spacetime is given by the direct sum  
\begin{equation}
R=\tilde{R}_d+\gamma g^{ik}g^{jl}\left(\hat{g}_{i k} \hat{g}_{j l}-\hat{g}_{i l} \hat{g}_{j k}\right)=\tilde{R}_d+\hat{R}_p.
\end{equation}
In general, compactifications of a given theory on direct product spaces produce incompatibilities in between these field equations. In fact, in the absence of the QTE Lagrangian, Einstein-Maxwell theory is clearly incompatible unless severe restrictions on the form of the electric field are imposed. To better observe these incompatibilities, it is convenient to compare the trace of the field equation on the brane and on the internal manifold. From (\ref{EOMQTE}) we get 
\begin{align}
\tilde{R}_d\left(1-\frac{d}{2}\right)+d\left(\Lambda-\frac{\hat{R}_p}{2}+\frac{1}{4p!}\hat{H}^2\right)-\left(\frac{1}{2}+4p!\beta_0 \hat{H}^2\right)\left(1-\frac{d}{4}\right)\tilde{F}^2&=0\nonumber\\
-p\frac{\tilde{R}_d}{2}+\left(\hat{R}_p\left(1-\frac{p}{2}\right)+p\Lambda-\frac{1}{4(p-1)!}\hat{H}^2\right)-p\left(p!\beta_0 \hat{H}^2 -\frac{1}{8}\right)\tilde{F}^2&=0
\end{align}
whose compatibility is subjected to the conditions 
\begin{align}
Q_m^2&=\frac{1}{8\beta_0}\frac{1}{(d-3)p!^2},\label{comp1}\\
\Lambda&=\frac{1}{2}(p-1)(d+p-2)\gamma-\frac{d-1}{4}Q_m^2 \label{comp2}.
\end{align}
Here, we have made explicit the fact that under (\ref{cuv}) and (\ref{fieldans}), $\hat{R}_p=\gamma p(p-1)$ and $\hat{H}^2=p!Q_m^2$, respectively. A few comments are in order. From (\ref{comp1}), it is direct to observe how the Lagrangian (\ref{QTE}) makes possible the compactification of the theory through the inclusion of the magnetic charge, which when properly fixed cures an incompatibility introduced by the presence of $\tilde{F}^2$.
Indeed, for any dimension $d$, in the absence of the $\beta_0$ term, the field equations are compatible only for $\tilde{F}^2=0$, a condition that cannot be accomplished by purely electric solutions. Hence, the inclusion of the QTE Lagrangian opens the road to compactify Einstein-Maxwell theory over $\mathcal{M}_D=\mathcal{M}_d\times\mathcal{K}^p$ generically, without restricting the spacetime dimension and neither the nature of the electromagnetic field living on the brane, with the three-dimensional brane case a  branch to be analyzed separately. 
The cosmological constant acquires two contributions, one coming from $\mathcal{K}^p$ through its curvature $\gamma$ and another due to the kinetic term for $\hat{H}_{i_1...i_p}$. 
Then, to have generic asymptotically flat solutions the internal manifold is forced to be flat while at the same time in (\ref{QTE}) we must disregard the standard kinetic term.\\
In the next section, we proceed to integrate the field equations (\ref{EOMQTE}) addressing the construction of Reissner-Nordstr\"om black $p$-branes in dimension $D=d+p$, where the internal manifold has a generic constant curvature $\gamma$.  

\subsection{Reissner-Nordstrom black string and $p$-branes}

After the viability of the compactification is ensured, the field equations become integrable, enabling the construction of simple cylindrically extended objects, black string and black $p$-branes. 
For the sake of simplicity, let us describe $\mathcal{M}_d$ by the following $d$-dimensional spherically symmetric spacetime,
\begin{equation}
ds^2_d=-F(r)dt^2+\frac{dr^2}{F(r)}+r^2d\Sigma^2_{d-2,K}, \label{brane}
\end{equation}
where $d\Sigma^2_{d-2,K}$ stands for a $(d-2)$-dimensional Euclidean manifold of normalized constant curvature $K=0,\pm1$\footnote{This is a simplicity assumption. Indeed, we might take any solution of the Einstein-Maxwell-$\Lambda$ equations.}. 
Integrating out the Einstein-Maxwell-$\Lambda$ equations on the brane, and including the compatibility conditions (\ref{comp1}) and  (\ref{comp2}), the following Reissner-Nordstr\"om black $p$-brane is obtained, 
\begin{equation}
ds_D^2=-F(r)dt^2+\frac{dr^2}{F(r)}+r^2d\Sigma^2_{d-2,K}+\frac{d \vec{z} \cdot d \vec{z}}{\left(1+\frac{\gamma}{4} \sum_{j=1}^{p} z_{j}^{2}\right)^{2}} \label{sol}
\end{equation}
with 
\begin{equation}
F(r)=\frac{r^2}{l_{eff}^2}-\frac{M}{r^{d-3}}+K+\frac{Q_e^2}{2(d-3)^2r^{2(d-3)}},
\end{equation}
and an effective AdS radius defined by
\begin{equation}
l_{eff}^{-2}:=-\frac{1}{16\beta_0}\frac{32\beta_0\Lambda(d-3)p!^2-(p-1)}{(d-1)(d-3)(d+p-2)p!^2}.
\end{equation}
We observe that the net effect of the magnetic field on the spacetime backreaction lies in to provide the brane black hole with an effective electric chargelike term,
\begin{equation}
\bar{Q}_{e}^{2}:= \frac{1}{2(d-3)^2}Q_e^2, \label{effcharge}
\end{equation}
which differs from the one of a standard Reissner-Nordstr\"om black hole $Q_e^{RN}:=\frac{Q_e^2}{2(d-2)(d-3)}$, $\bar{Q}_{e}<Q_e^{RN}$. This can be seen in form (\ref{EOMQTE}): the field equations on the brane possess a Maxwell-Faraday stress tensor which is modulated by an effective coupling $(1+8p!\beta_0H^2)$ which departs from the standard value due to the presence of the magnetic forms. In consequence, the backreaction on the metric has the same falloff of the Reissner-Nordstr\"om solution but with an effective physical charge, as the direct computation of the Gauss law reveals. \\
Solution (\ref{sol}) represents the most general homogenous black $p$-brane extension of the Reissner-Nordstr\"om black hole.
No restrictions prevent us from taking the $p=1$ ($\gamma=0$) case and, in consequence, to construct the simplest black string generalization of the Reissner-Nordstr\"om black hole. Avoiding the inclusion of the kinetic term for $\hat{H}_{i_1...i_p}$ and the bare cosmological constant $\Lambda$, the solution yields
\begin{equation}
ds^2=-\left(K-\frac{M}{r}+\frac{Q^2}{2r^{2}}\right)dt^2+\frac{dr^2}{\left(K-\frac{M}{r}+\frac{Q^2}{2r^{2}}\right)}+r^2d\Sigma^2_{2,K}+dz^2.
\end{equation}
Notice that, although we have obtained this black string by considering a general $p$-form over the internal manifold and then taking the $p=1$ case, the same result can be achieved by starting immediately with $K^p=\mathbb{R}$ or $\mathcal{S}^1$ and dressing the internal space with a scalar field depending linearly on the flat coordinate. For such a case, only when going to the $(d+p)$-dimensional case, a difference is observed. As a matter of fact, when dressing the internal manifold with scalar fields, it is necessary to include one scalar along each extended direction,\footnote{This case would correspond to an internal manifold which is divided into $p$ subinternal spaces of dimension one. In the next section, we face the splitting of the internal manifold in full generality, showing the restrictions that the number of subdivisions and the corresponding dimension $d$ must accomplish in order for the solutions to exist.} and in consequence, $p$ scalars are required. This translates into an effective electric charge, 
\begin{equation}
\bar{Q}_{e}^{2}:= \frac{1}{2(d-3)(d-p-2)}Q_e^2. \label{scalarcase}
\end{equation}
Indeed, when considering one $p$-form instead of $p$ scalar fields, $p$ goes to one, and (\ref{effcharge}) is recovered. 

\subsection{Compactifications of Einstein-Maxwell theory on factorized internal manifolds}

From the compatibility condition (\ref{comp1}) we observe that, beyond the logarithmic branch introduced by the electric field in three dimensions, the three-dimensional case is pathological, and no black string solutions exist. However, we observe that for the scalar case in which the flat internal manifold $\mathcal{K}^p$ is subdivided in $p$ flat internal manifolds, the compatibility condition changes into
\begin{equation}
Q_{m}^2=\frac{1}{8\beta_0}\frac{1}{(d-p-2)},
\end{equation}
which takes into account the number of subdivisions, $p$ in this case, of the internal manifold. Note that in this case the number of subdivisions corresponds with the number of scalar fields.  In (\ref{scalarcase}), the three-dimensional case is not prohibited unless the internal manifold is one dimensional, i.e., no appearance of a black string extension of the charged BTZ black hole. We observe that for factorized internal manifolds there is a critical dimension in which, for specific values of the number of subdivisions, the theory does not support compactifications, therefore the existence of black $p$-branes. \\ 
Let us consider an internal manifold $\mathcal{K}^p$ given by a direct product of internal manifolds of lower dimensionality,
\begin{equation}
\mathcal{K}^p=\mathcal{K}^{p_1}\times\mathcal{K}^{p_2}\times\cdots\mathcal{K}^{p_n},
\end{equation}
where the sum of the dimensions of all the $n$ subinternal manifolds $\mathcal{K}^{p_n}$ must correspond to the dimension of the original internal manifold $\mathcal{K}^{p}$, $\sum_n p_n=p$. 
It must be noticed that the dimensions $p_n$ are arbitrary. Thus, the number of possible subdivisions and combinations of them that reproduce the original internal manifold corresponds to the number of different manners in which a natural number can be written as the sum of an arbitrary number of lower naturals, namely, the partitions of $p$, which are asymptotically given by the Hardy-Ramanujan formulas. Here, we care about the value $n$ that gives the number of submanifolds that are considered into the original splitting and not about the number of possible combinations that provide an internal manifold of an equivalent dimension. This is due to the fact that the distribution of the subinternal manifolds is irrelevant in terms of the backreaction of the solutions. 
To proceed with the compactification and in order to 
avoid any undesired relation in between the parameters of our theory, we dress each internal submanifold with an interaction
\begin{equation}
\mathcal{L}_{QTE}^j=-2p_j!\beta_{0,j} F_{A_1A_2}F^{A_1A_2}H_{(j),B_1...B_{p_j}}H_{(j)}^{B_1...B_{p_j}}, 
\end{equation}
where, again, we consider the corresponding standard kinetic term for each of the magnetic forms. 
%-\frac{1}{2p_j!}H_{(j),A_1\dots A_{p_j}}H_{(j)}^{A_1\dots A_{p_j}}
Then, Einstein-Maxwell theory is complemented with 
\begin{equation}
L_{QTE=}\sum_{j=1}^n\left(-\frac{1}{2p_j!}H_{(j),A_1\dots A_{p_j}}H_{(j)}^{A_1\dots A_{p_j}}+\mathcal{L}_{QTE}^j\right). 
\end{equation}
Each of the $p_n$-form fields will be proportional to the volume form of the corresponding $\mathcal{K}^{p_n}$ subspace. Notice that each interaction is controlled by its own coupling constant $\beta_{0,n}$ and that $n$ only labels the number of subinternal manifolds.
The Ricci scalar of the whole spacetime and the square of the magnetic $p$-form field strength are now given by 
\begin{align}
R&=\tilde{R}_d+\sum_{j=1}^{n}\gamma_jp_j(p_j-1),\\
\hat{H}_{(n)}^2&=p_n! Q_{m_n}^2,
\end{align}
with $\gamma_j$ defining the curvature radius of the subinternal manifold $\mathcal{K}^{p_j}$. 
Due to the nature of our product spacetime the corresponding field equations can be projected along the spacetime manifold $\mathcal{M}_d$ and along each subinternal manifold $\mathcal{K}^{p_n}$. The trace of the field equations on $\mathcal{M}_d$ is given by 
\begin{equation}
\tilde{R}_d=\frac{2d}{(d-2)}\Lambda^{d}_{eff}+e_{eff}^d\left(\frac{d-4}{d-2}\right)\frac{\tilde{F}^2}{2},  \label{multitrace}
\end{equation}
where
\begin{align}
\Lambda^{d}_{eff}&=\sum_{j=1}^n\left(\frac{1}{4p_j!}\hat{H}_{(j)}^2-\frac{1}{2}\hat{R}_{p_j}\right)+\Lambda\label{multicond1},\\
e_{eff}^d&=\frac{1}{2}+4\sum_{j=1}^np_j!\beta_{0,j} \hat{H}_{(j)}^2, \label{multicond2}
\end{align}
represent the effective contributions of the magnetic forms on the brane cosmological constant and on the effective Maxwell-Faraday stress tensor. \\
For the field equations on the whole spacetime to be compatible, Eqs. (\ref{multitrace}) along with (\ref{multicond1}) and (\ref{multicond2}) must be compatible with the field equations projected on all subinternal manifolds $\mathcal{K}^{p_n}$. This defines the following $2n$-dimensional system of equations:
\begin{align}
\Lambda^{p_i}_{eff}-\frac{p_id}{d-2}\Lambda^{d}_{eff}&=0, \label{eq1}\\
e_{eff}^{p_i}-\frac{e_{eff}^d}{4}\left(\frac{d-4}{d-2}\right)&=0 \label{eq2},
\end{align}
where on each subinternal manifold has been defined an effective cosmological constant and an effective contribution of the $p_n$-form fields on the electromagnetic stress tensor,
\begin{align}
\Lambda^{p_i}_{eff}&=\hat{R}_{p_i}\left(1-\frac{p_i}{2}\right)-\frac{p_i}{2}\left(\sum_{j=1}^n\hat{R}_{p_j}\right)+p_i\Lambda-\frac{\hat{H}_i^2}{4(p_i-1)!}+p_i\left(\sum_{j=1}^n\frac{1}{4p_j!}\hat{H}_j^2\right), \hspace{0.3cm} j\neq i \label{condbrane1}\\
e_{eff}^{p_i}&=\frac{1}{8}+\left(\sum_{j=1}^np_j!\beta_{0,j}\hat{H}_j^2-2p_j!\beta_{0,i}\hat{H}_i^2\right).\label{condbrane2}
\end{align}
On the left-hand side, $p_i$ labels the corresponding subinternal manifold in which the field equations have been projected. For the right-hand side, $p_i$ is the dimension of the corresponding $\mathcal{K}^{p_i}$ on which the $\hat{H}_{(i)}$ field has been distributed.
There is no summation on $i$, and it must be noticed that in Eq. (\ref{condbrane1}) the sums are carried out on all values of $j$ except for the one defined by $i$. Equation (\ref{eq1}) defines the compatibility relations that all contributions to the effective cosmological constant must satisfy, while Eq. (\ref{eq2}) defines the compatibility conditions that all terms proportional to the Maxwell field must accomplish. Equations (\ref{eq1}) and (\ref{eq2}) constitute a nonhomogenous linear system for $Q_{m_n}$ and $\gamma_n$.\\
To analyze the existence of solutions, it is useful to write the system in the following matrix form: $A_{2n}\vec{X}=\vec{B}$, with $A$ a coefficient matrix of order $2n$ and $B$ a $2n$-dimensional column vector representing the inhomogeneous contributions in the system. 
For the system to have a solution, the rank of $A$ must be equal to the rank of the enlarged matrix $\bar{A}$, which is obtained by adding into $A$ one more column given by the vector $B$.
If this is ensured, a unique solutions exist when the rank of $A$ is equal to its order; otherwise, if the rank of $A$ is less, infinite solutions emerge. \\
In our case, for arbitrary values of the dimension $d$ and the dimensions $p_n$ of each subinternal manifold, the system possesses a single solution given by  
\footnotesize
\begin{align}
\gamma_i&=\frac{1}{(p_i-1)(d+p-2)}\left[2\Lambda-\frac{1}{2}\sum_{j=1}^n(p_j-1)Q_{m_j}^2+\frac{1}{2}(p-p_i)Q_{m_i}^2+\frac{1}{2}(d-1)Q_{m_i}^2\right], j\neq i \label{congen1}\\
Q_{m_i}^2&=\frac{1}{8\beta_{0,i}p_i!^2(d-n-2)}\label{congen2}.
\end{align}
\normalsize
We observe that for a solution to emerge, with one or multiple internal manifolds, it is mandatory that the dimension of the spacetime $\mathcal{M}_d$ and the number of internal manifolds satisfy $d-n-2\neq 0$.
This condition is precisely violated for the three-dimensional case in which only one internal manifold is considered. For such a case, the rank of $A$ is different from the rank of the enhanced matrix $\bar{A}$, and therefore, no solution exists, the same problem that the four-dimensional case encounters when considering two internal manifolds, no matter their respective dimensions. We also note that, when the internal manifold is divided into subinternal manifolds of dimension one, the dimension of the original internal manifold and the number of subdivisions are equal. This is the case when $p$ scalar fields are distributed along $p$ flat extended directions.
Finally, in the limit for which $n=1$, and in consequence only one charge appears, conditions (\ref{congen1}) and (\ref{congen2}) precisely reduce to (\ref{comp1}) and (\ref{comp2}).

\section{Compactification of Einstein-Maxwell-Lovelock theory}

Black strings are by definition higher-dimensional objects, and in consequence, it is appealing to understand whether or not they belong to the black hole spectrum of Lovelock theory. 
It has been already shown that in order to construct black strings and $p$-branes in Lovelock gravity, it is mandatory to include nonminimal couplings in between curvature tensors and magnetic $p$-forms \cite{Cisterna:2020kde}, this along the lines of the theory defined by (\ref{eq:general-action-matter}). In essence, the inclusion of higher curvature terms naturally introduces incompatibilities between the projected field equations. Then, the presence of new curvature tensors, this time modulated by their couplings to $p$-forms, opens the road to cure these obstructions. Here we further extend these findings to consider the inclusion of the Maxwell field. 

\subsection{Black strings in Einstein-Maxwell-Gauss-Bonnet theory}

We start by considering Einstein-Maxwell-Gauss-Bonnet gravity
\begin{equation}
S_{EMGB}=\int{\left(R-2\Lambda-\frac{1}{4}F^2+\alpha_2\mathcal{L}^{(2)}\right)}\sqrt{-g}d^{d+p}x, \label{EMGB}
\end{equation}
with a brane dimension higher than four in order to ensure a nontrivial contribution of the Gauss-Bonnet term into the dynamic of the theory. 
After the field equations are projected, two types of incompatibilities might arise. The first one is generated by the presence of the Gauss-Bonnet density, which is quadratic in the curvature tensor and, once projected, introduces incompatibilities with the Einstein term. The second type of incompatibility comes directly from the inclusion of the Maxwell Lagrangian.
Despite these generic features, it is important to note that for a pure Gauss-Bonnet theory (no Einstein term) supplemented with the Maxwell action the compactification is trivial and in fact no need of any extra matter fields is required \cite{Giacomini:2018sho}. This result is extended beyond the domain of pure Gauss-Bonnet theory and states that any pure Lovelock theory of order $k$ supports the existence of homogeneously charged black $p$-branes as long as the Maxwell field living on the brane is given by a $p$-form field of order $p=k$.
In this way, pure Gauss-Bonnet theory naturally contains electrically charged black strings, i.e., $k=p=2$. \\
Considering now the presence of the Einstein term, %i.e. the whole Lovelock series up to a given order $k$,
we note that only the incompatibilities coming from the curvature invariants of different order must be fixed. At this point, to construct charged black $p$-branes in Einstein-Gauss-Bonnet theory (EGB) it is enough to consider the approach developed in \cite{Cisterna:2020kde} with no extra interactions of the QTE form (\ref{QTE}) we previously considered. 
Thus, we start complementing (\ref{EMGB}) with $\mathcal{L}^{(1,1)}$ and its corresponding kinetic term
\begin{equation}
S^{(1,1)}=\int{\left(-\frac{1}{2p!}\hat{H}_{(1)}^2+\frac{\beta_1}{2p!}\delta^{A_1A_2C_1...C_p}_{B_1B_2D_1...D_p}R^{B_1B_2}{}_{A_1A_2}H^{D_1...D_p}{}H_{C_1...C_p}\right)}\sqrt{-g}d^{d+p}x.
\end{equation}
This interaction naturally cures the incompatibilities in between the projected Gauss-Bonnet density and the Ricci scalar. The corresponding field equations are given in the Appendix where the explicit compatibility conditions are defined by relations (\ref{cc1}) and (\ref{cc2}). As an example, we explicitly construct a seven-dimensional (A)dS-charged Boulware-Deser black 2-brane composed by the direct product of a charged Boulware-Deser black hole in dimension five and a two-dimensional space of constant curvature, i.e., $d=5$ and $p=2$. The solution is given by
\begin{equation}
ds^2=-F(r)dt^2+\frac{dr^2}{F(r)}+r^2d\Sigma^2_{K,3}+\frac{dz_1^2+dz_2^2}{1+\frac{\gamma}{4}(z_1^2+z_2^2)^2},
\end{equation}
where
\begin{equation}
F(r)=K+\frac{\hat{\xi}_1}{4\alpha_2}r^2\left(1\pm\sqrt{1+\frac{4}{3}\frac{\alpha_2}{\hat{\xi}_1^2}\Lambda_{brane}+\frac{8\alpha_2}{3\hat{\xi}_1^2}\frac{M}{r^4}-\frac{2\alpha_2}{3\hat{\xi}_1^2}\frac{Q_e}{r^6}}\right).
\end{equation}
The compatibility conditions have been introduced in terms of $Q_{m_1}^2$ and the bare cosmological constant $\Lambda$, fixing the effective coupling $\hat{\xi}_1$ and the brane cosmological constant to be
\begin{align}
\hat{\xi}_1&=\frac{1}{2}+\alpha_2\gamma,\\
\Lambda_{brane}&=\frac{1}{32\beta_1}(1+2\gamma(3\alpha_2+4\beta_1)).
\end{align}
This solution is obtained from the effective Wheeler polynomial, 
\begin{equation}
-\frac{1}{2}\hat{\xi}_1(K-F(r))-\frac{1}{r^2}\alpha_2(K-F(r))^2+\frac{1}{3}\frac{M}{r^3}+\frac{1}{12}\Lambda_{brane}r^2-\frac{1}{24}\frac{Q_e^2}{r^4}=0,
\end{equation}
where $M$ and $Q_e$ are integration constants ultimately related with the physical mass and electric charge \cite{Cai:2001dz,Cai:2003gr,Kofinas:2007ns}. 
Note that the $p$-form fields distributed on the internal manifold affect the solutions through the effective Newton constant coupling $\hat{\xi}_1$, changing the numerical factors in the values of the mass and electric charge, this beyond the usual modifications introduced by the higher curvature terms of Lovelock gravity.  
In order for these conserved quantities to be well defined, $\hat{\xi}_1$ must be positive, which is generically ensured when compactifying over internal manifolds of positive constant curvature. Consequently, the coupling $\beta_1$ takes negative values in order for $Q_{m_1}$ to be real. On the other hand, for negatives values of $\gamma$, the positivity of $\hat{\xi}_1$ can be maintained by constraining the internal manifold curvature radius in terms of the Gauss-Bonnet coupling. For $\beta_1>0$ the internal manifold curvature is bounded from above in terms of $\alpha_2$, while for $\beta_1<0$ it is constrained to take values on a restricted interval defined by $\alpha_2$.
As in standard Einstein-Gauss-Bonnet black holes, the negative branch connects asymptotically with the corresponding GR solutions, and it is regarded as the physical branch. This black brane possesses a Reissner-Nordstr\"om-like limit contrary to the case exposed in \cite{Giacomini:2018sho}, and it simply represents the charged extension of the one presented in \cite{Cisterna:2020kde} whose existence is provided by the same compatibility relations presented here.\\
It is important to stress that these solutions represent homogenous black branes that precisely solves Einstein-Gauss-Bonnet equations on the brane, now with shifted coupling constants. This makes a difference with a related family of solutions previously found in \cite{Maeda:2006iw,Maeda:2006hj,Cai:2009de}. In there, the field equations on the brane are automatically satisfied by fixing the couplings in such a way that those field equations vanish identically. Then, the metric is integrated by solving the field equations on the internal manifold, which basically translate into solve a trace of the form $\kappa_1 R+\kappa_2\mathcal{L}^{(2)}+\kappa_3 g_{\mu\nu}=0$, with $\kappa_i$ representing the corresponding couplings up to some numerical factors. These solutions possess an EGB shape; nevertheless, due to the fact that they solve the internal manifold field equations, they have a different falloff in the mass and charge terms and do not match the Reissner-Nordstr\"om limit for large values of $r$.
The causal structure of our compactified black holes is qualitatively the same than the standard black holes of EGB theory \cite{Charmousis:2008kc,Garraffo:2008hu}, with an interesting difference when analyzing the case in which the theory possesses one degenerated vacuum. These solutions were exhaustively analyzed in \cite{Crisostomo:2000bb}. 
Such a case emerges for a precise tuning between the cosmological constant and the Gauss-Bonnet coupling, defining what is known as Chern-Simon theories, see \cite{Zanelli:2005sa} and references therein. The structure of the solutions is markedly different from black holes with nondegenerated vacua, and it even allows for a rotating black hole \cite{Anabalon:2009kq}. Here, we emphasis that in our solutions such a point do also exists; nevertheless, the presence of the $p$-form fields dissolves the undesirable tuning of the cosmological constant and the Gauss-Bonnet coupling by introducing the magnetic charges, which are generic integration constants. For our model the Chern-Simons point reads
\begin{equation}
0=1+\frac{4}{3}\frac{\alpha_2}{\hat{\xi}_1^2}\Lambda_{brane}\rightarrow 24\beta_1\alpha_2^2\gamma^2+2\alpha_2(16\beta_1+3\alpha_2)\gamma+6\beta_1+\alpha_2=0, 
\end{equation} 
defining an open set of the parameter space in which the coupling constants are free due to the presence of $\gamma$.

\subsection{Black strings in Einstein-Maxwell-Lovelock theory: The cubic case}
 
Moving to the next order in Lovelock theory, namely, to the case in which the cubic Lovelock density is included, the compactification requires more effort. According to \cite{Giacomini:2018sho}, a pure  cubic Lovelock theory will support trivial compactifications as long as the Maxwell field is given by a magnetic 3-form. Hence, the black hole on the brane cannot be electrically charged. Moreover, due to the fact that the cubic density contributes to the field equations for $d>6$, the existence of these charged black $p$-branes is also subjected to the dimensionality of the black hole horizon, six in the minimalistic case, and in accordance to this, at least two $3$-form fields need to be consider. \\
For a Lovelock series of order three, in the presence of the Einstein and Gauss-Bonnet terms, we observe no simplification in contrast to the previous case. 
In order to construct electrically charged black $p$-branes, we need to take into account the incompatibilities produced by the curvature quantities of different order and the incompatibilities added by the electric Maxwell field. 
Consequently, in generic Lovelock theories of order $k>2$, we need to combine the procedure of \cite{Cisterna:2020kde} with a term of the QTE form (\ref{QTE}).
For Einstein-Maxwell-cubic Lovelock theory, we consider the following action principle,
\small 
\begin{equation}
S_{EMCL}=\int{\left(R-2\Lambda-\frac{F^2}{4}+\alpha_2\mathcal{L}^{(2)}+\alpha_3\mathcal{L}^{(3)}-\frac{H_{(1)}^2}{2p!}-\frac{H_{(2)}^2}{2p!}-\frac{H_{(3)}^2}{2p!}+\mathcal{L}^{(1,1)}_{(1)}+\mathcal{L}^{(2,1)}_{(2)}+\mathcal{L}^{QTE}_{(3)}\right)}\sqrt{-g}d^{d+p}x, \label{cubic}
\end{equation}
\normalsize
where 
\begin{equation}
\mathcal{L}^{(2,1)}=\frac{\beta_2}{4p!}\delta^{A_1A_2A_3A_4C_1...C_p}_{B_1B_2B_3B_4D_1...D_p}R^{B_1B_2}{}_{A_1A_2}R^{B_3B_4}{}_{A_3A_4}H^{D_1...D_p}{}H_{C_1...C_p}.
\end{equation}
This term corresponds to the next curvature order contained in (\ref{eq:general-action-matter}), and it provides compatibility in the presence of the cubic Lovelock density.
We recall that the cubic density contributes to the field equations for $d>6$. Note that we have also included the kinetic terms associated with $\mathcal{L}^{(1,1)}_{(1)}$ and $\mathcal{L}^{(2,1)}_{(2)}$ and that each of the interactions is given by different magnetic $p$-forms, ${H}_{(1)}$, ${H}_{(2)}$, and ${H}_{(3)}$ corresponding to $\mathcal{L}^{(1,1)}_{(1)}$, $\mathcal{L}^{(2,1)}_{(2)}$, and $\mathcal{L}^{QTE}_{(3)}$, respectively.\\
The field equations are given in a compact form in the Appendix, relations 
(\ref{eqmc1}) and (\ref{eqmc2}).\\
For the sake of concreteness, here we provide 
the explicit construction of a ten-dimensional black 2-brane which is composed of an eight-dimensional charged Boulware-Deser black hole and a two-dimensional space of constant curvature, i.e., $d=8$ and $p=2$. By plugging these dimensions on the corresponding compatibility relations (\ref{CCC1}) and (\ref{CCC2}), the magnetic charges are fixed to be 
\begin{align}
Q_{m_1}^2 &= -\frac{3}{2}\frac{\alpha_2}{\beta_1}\gamma-\frac{1}{4\beta_1},\label{cccc1}\\
Q_{m_2}^2 &= -\frac{2\alpha_3}{\beta_2}\gamma-\frac{1}{6}\frac{\alpha_2}{\beta_2},\label{cccc2}\\
Q_{m_3}^2 &=-\frac{1}{96\beta_0} \label{cccc3},
\end{align}
while the bare cosmological constant is related with the curvature radius of the internal manifold through the relation 
\begin{align}
\Lambda= \left(\frac{5}{8}\frac{\alpha_2}{\beta_1}+\frac{5}{6}\frac{\alpha_3}{\beta_2}+\frac{4}{3}\right)\gamma+\frac{5}{72}\frac{\alpha_2}{\beta_2}+\frac{5}{48\beta_1}+\frac{5}{1152\beta_0}. \label{cccc4}
\end{align}
The field equations are then integrable and provide a charged black $2$-brane of the form 
\begin{equation}
ds^2=-F(r)dt^2+\frac{dr^2}{F(r)}+r^2d\Sigma_{k,6}^2+\frac{dz_1^2+dz_2^2}{(1+\frac{\gamma}{4}(z_1^2+z_2^2))^2},
\end{equation}
with the lapse function given by the real solution of the effective third-degree Wheeler polynomial,
\begin{equation}
-3r^5(K-F(r))\hat{\xi}_1-60r^3(K-F(r))^2\hat{\xi}_2-360r(K-F(r))^3\alpha_3-\frac{\hat{\xi}_4}{20}\frac{Q_e^2}{r^5}+\frac{1}{7}\Lambda_{brane}r^7+M=0.
\end{equation}
After applying the corresponding compatibility conditions (\ref{cccc1})-(\ref{cccc3}) and (\ref{cccc4}), the effective couplings are 
\begin{align}
\hat{\xi}_1 &= \frac{1}{2}+\alpha_2\gamma,\hspace{0.3cm} \hat{\xi}_2 = \frac{2}{3}\alpha_2+2\alpha_3\gamma,\hspace{0.3cm} \hat{\xi}_4= \frac{2}{3}\\
\Lambda_{brane} &= \frac{1}{12\beta_1\beta_2}(4\alpha_3\beta_1+4\beta_1\beta_2+3\alpha_2\beta_2)\gamma+\frac{1}{576\beta_1\beta_2\beta_0}(24\beta_2\beta_0+\beta_1\beta_2+16\alpha_2\beta_1\beta_0).
\end{align}
In this case, note that $\beta_0$ must be negative; otherwise, $Q_{m_3}$ becomes complex. Moreover, from the effective definitions of $\hat{\xi}_1$ and $\hat{\xi}_2$, the geometric terms of the Wheeler polynomial will always possess the same sign for $\gamma$ positive; alternatively, for negative $\gamma$, the coupling constants are bounded accordingly, as in the EGB case. In addition, the signs of $\beta_1$ and $\beta_2$ must combine in such a way that all magnetic charges are real. Finally, to obtain generic asymptotically flat solutions, it is needed to kill the bare cosmological constant, consider a flat internal manifold, and to avoid the presence of the corresponding $p$-form kinetic terms.

\section{Conclusions}

In this paper, we have addressed the construction of charged homogenous black strings and black $p$-branes in Einstein-Maxwell and Einstein-Maxwell-Lovelock theories. In order to achieve these constructions, we have first shown how to compactify these theories on direct product spaces of the form $\mathcal{M}_D=\mathcal{M}_d\times\mathcal{K}^p$. It was required that the internal manifold be dressed by a specific flux of $p$-forms proportional to the corresponding volume form and with a dynamic governed by suitable nonminimal couplings with curvature and electromagnetic tensors. Such interactions are naturally provided by the theory defined in (\ref{eq:general-action-matter}). Our solutions represent homogenous black string/$p$-brane extensions of the Reissner-Nordstr\"om and charged Boulware-Deser black holes, with or without a cosmological constant. Moreover, the extra dimensions are encoded on an internal manifold of arbitrary constant curvature. We have also extended our findings to the case in which several internal manifolds are included, finding a relation in between the brane dimension and the number of internal manifolds such that the compactifications are performable. \\
As a matter of fact, in \cite{Cisterna:2019scr}, it has been demonstrated that homogenous black strings in AdS \cite{Cisterna:2017qrb} are perturbatively stable at first order when considering generic perturbations of the metric and the scalar field. The scalar acts in such a way that stabilizes the extended object. On these lines, it would be appealing to explore the stability of the homogenous Reissner-Norstr\"om black strings we propose here. For the case of Maxwell-Lovelock theories, we observe that compactifications influence the black hole on the brane through an effective Wheeler polynomial in which all coefficients acquire corrections from the magnetic $p$-forms. These corrections translate into effective thermodynamical quantities, mass and charge. It would be interesting to understand how these corrections express themselves when verifying the first law of thermodynamics or how they might influence the existence of phase transitions. Moreover, we have observed that for theories in which there is a single degenerated vacua the usual tuning in between the cosmological constant and, for example, the Gauss-Bonnet coupling for the quadratic case dissolves due to the presence of the magnetic charges. Along these lines, the properties of dimensionally continued black holes \cite{Banados:1993ur,Crisostomo:2000bb} might be revisited or even the structure of the rotating solution found in \cite{Anabalon:2009kq}.
Finally, another direction to explore will be to investigate how to couple higher dimensional p-forms with other matter sources, in such a way that theories with different matter fields on the brane are compactifiable when considering compactifications on direct product spacetimes or to consider the inclusion of other geometric objects as, for example, the presence of a nontrivial torsion \cite{Castillo-Felisola:2016kpe}. It seems that, for the particular case in which the initial theory enjoys conformal invariance, the construction of homogenous extended objects is particularly favored.

\section{Acknowledgments}

We would like to thank Cristian Erices for collaboration on early stages of this work and Jos\'e Barrientos, Crist\'obal Corral, Hideki Maeda, and Julio Oliva for useful comments and remarks. We especially appreciate discussions with Claudio Gallegos and Konstantinos Pallikaris regarding the structures discussed in the Sec. III-B. 
This work is partially funded by Beca Chile de Postdoctorado Grant No. 74200012, FONDECYT Grant No. 1210500 and Becas ANID de Doctorado Nacional (2017 and 2021) Grants No. 21171394 and No. 21212393.

\section{Appendix: Decompositions and field equations in Einstein-Maxwell-cubic Lovelock theory}

Spacetimes of the form $\mathcal{M}_D=\mathcal{M}_d\times\mathcal{K}^p$ simplify substantially the form of the Riemann tensor, which is given by 
\begin{equation}
R^{A_1A_2}{}_{B_1B_2}=\tilde{R}^{\mu_1\mu_2}{}_{\nu_1\nu_2}\delta^{\nu_1}_{B_1}\delta^{\nu_2}_{B_2}\delta^{A_1}_{\mu_1}\delta^{A_2}_{\mu_2}+\hat{R}^{ij}{}_{kl}\delta^{k}_{B_1}\delta^{l}_{B_2}\delta^{A_1}_{i}\delta^{A_2}_{j},
\end{equation}
where greek indices denote quantities on the brane, and latin indices denote quantities on the internal manifold. According to this, here we provide the explicit decomposition of all quantities used along this work.\\
\\
Lagrangians:
\begin{equation}
  \begin{split}
   R&=\tilde{R}_d+\hat{R}_p\\
    \mathcal{L}^2&=\tilde{\mathcal{L}}^2_{d}+\hat{\mathcal{L}}^2_{p}+2\tilde{R}_{d}\hat{R}_{p}\\
 \mathcal{L}^3&=\tilde{\mathcal{L}}^3_d+3\tilde{\mathcal{L}}^2_d\hat{R}_p+3\hat{\mathcal{L}}^2_p\tilde{R}_d+\hat{\mathcal{L}}^3_p
  \end{split}
\quad\quad
  \begin{split}
   \mathcal{L}^{(1,1)}_{(1)}&=\tilde{R}_d\hat{H}^2_{(1)}\\
   \mathcal{L}^{(2,1)}_{(2)}&=\tilde{\mathcal{L}}^2_d\hat{H}^2_{(2)}\\
   \mathcal{L}^{QTE}_{(3)}&=-\frac{1}{2p!}\hat{H}^2_{(3)}-2p!\beta_0\tilde{F}^2\hat{H}^2_{(3)}.
  \end{split}
\end{equation}
Relevant tensors:
\small
\begin{align}
G_{AB}&=(\tilde{G}_{\mu \nu}^d-\frac{1}{2}\tilde{g}_{\mu \nu}\hat{R}_{p})\delta_{A}^{\mu}\delta_{B}^{\nu} + (\hat{G}_{ij}^p -\frac{1}{2}\hat{g}_{ij}\tilde{R}_{d})\delta_{A}^{i}\delta_{B}^{j}  \\
\mathcal{H}_{AB}&=(\tilde{\mathcal{H}}_{\mu\nu}^d+2\tilde{G}_{\mu \nu}^d\hat{R}_p-\frac{1}{2}\tilde{g}_{\mu\nu}\hat{\mathcal{L}}^2_p)\delta_{A}^{\mu}\delta_{B}^{\nu}+(\hat{\mathcal{H}}_{ij}^p+2\hat{G}_{ij}^p\tilde{R}_d-\frac{1}{2}\hat{g}_{ij}\tilde{\mathcal{L}}^2_d)\delta_{A}^{i}\delta_{B}^{j}\\
\mathcal{E}_{AB}&=(\tilde{\mathcal{E}}_{\mu\nu}^d+3\tilde{\mathcal{H}}_{\mu\nu}^d\hat{R}_p+3\tilde{G}_{\mu \nu}^d\hat{\mathcal{L}}^2_p-\frac{\tilde{g}_{\mu\nu}}{2}\hat{\mathcal{L}}_p^3)\delta_{A}^{\mu}\delta_{B}^{\nu}+(\hat{\mathcal{E}}_{ij}^p+3\hat{\mathcal{H}}_{ij}^p\tilde{R}_d+3\hat{G}_{ij}^p\tilde{\mathcal{L}}^2_d-\frac{\hat{g}_{ij}}{2}\tilde{\mathcal{L}}_d^3)\delta_{A}^{i}\delta_{B}^{j}\\
T_{AB}^{(1,1)}&=-\tilde{G}_{\mu\nu}^d\hat{H}^2_{(1)}\delta_{A}^{\mu}\delta_{B}^{\nu}-(p\tilde{R}_d\hat{H}_{(1)i}{}^{l_2...l_p}\hat{H}_{(1)jl_2...l_p}-\frac{1}{2}\hat{g}_{ij}\tilde{R}_d\hat{H}^2_{(1)})\delta_{A}^{i}\delta_{B}^{j}\\
T_{AB}^{(2,1)}&=-\tilde{\mathcal{H}}_{\mu\nu}^d\hat{H}^2_{(2)}\delta_{A}^{\mu}\delta_{B}^{\nu}-(p\tilde{\mathcal{L}}^2_d\hat{H}_{(2)i}{}^{l_2...l_p}\hat{H}_{(2)jl_2...l_p}-\frac{1}{2}\hat{g}_{ij}\tilde{\mathcal{L}}^2_d\hat{H}^2_{(2)})\delta_{A}^{i}\delta_{B}^{j}\\
T_{AB}^{QTE}&=2p!(2\tilde{F}_{\mu\lambda}\tilde{F}_\nu{}^\lambda\hat{H}^2_{(3)}-\frac{1}{2}\tilde{F}^2\hat{H}^2_{(3)})\delta_{A}^{\mu}\delta_{B}^{\nu}+2p!(p\hat{H}_{(3)ii_2...i_p}\hat{H}_{(3)j}{}^{i_2...i_p}\tilde{F}^2-\frac{1}{2}\tilde{F}^2\hat{H}^2_{(3)})\delta_{A}^{i}\delta_{B}^{j}.\
\end{align}
\normalsize
For the sake of compactness, we have not included the energy-momentum tensors associated to the kinetic term of each $p$-form field $\hat{H}_{(i)i_1...i_p}$; these are direct to compute. \\
For the Einstein-Maxwell-cubic Lovelock theory considered in Sec. IV-B, the field equations can be written in the following compact form:
\begin{align}
\hat{\xi}_1 \tilde{G}_{\mu\nu}^d+\Lambda_{brane}\tilde{g}_{\mu\nu}+\hat{\xi}_2\tilde{\mathcal{H}}_{\mu\nu}^d+\alpha_3\tilde{\mathcal{E}}_{\mu\nu}^d-\hat{\xi}_{em}\left(\tilde{F}_{\mu\lambda}\tilde{F}_{\nu}{}^{\lambda}-\frac{1}{4}\tilde{g}_{\mu\nu}\tilde{F}^2\right)&=0\label{eqmc1}\\
\hat{ \epsilon}^1_{ij}\tilde{R}_d + \hat{\epsilon}^0_{ij} + \hat{\epsilon}^2_{ij}\tilde{\mathcal{L}}^2_d-\frac{1}{2}\alpha_3\hat{g}_{ij}\tilde{\mathcal{L}}^3_d-\hat{\epsilon}^{em}_{ij}\tilde{F}^2&=0\label{eqmc2},
\end{align}
where we have defined the auxiliary quantities 
\begin{align}
\Lambda_{brane}&=-\frac{\hat{R}_p}{2}+\Lambda-\frac{\alpha_2}{2}\hat{\mathcal{L}}^2_p-\frac{\alpha_3}{2}\hat{\mathcal{L}}^3_p+\frac{1}{4p!}(\hat{H}_{(1)}^2+\hat{H}_{(2)}^2+\hat{H}_{(3)}^2)\\
\hat{\xi}_1&=1+2\alpha_2\hat{R}_p+3\alpha_3\hat{\mathcal{L}}^2_p+\beta_1\hat{H}_{(1)}^2\\
\hat{\xi}_2&=\alpha_2+3\alpha_3\hat{R}_p+\beta_2\hat{H}_{(2)}^2\\
\hat{\xi}_{em}&=\frac{1}{2}+4\beta_0p!\hat{H}_{(3)}^2
\end{align}
on the brane, and
\begin{align}
\hat{\epsilon}^0_{ij}&=\hat{G}_{ij}^p+\hat{g}_{ij}\Lambda+\alpha_2\hat{\mathcal{H}}_{ij}^p+\alpha_3\hat{\mathcal{E}}_{ij}^p-\sum_{m=1}^3\left(\frac{1}{2(p-1)!}\hat{H}_{(m)ii_2...i_p}\hat{H}_{(m)j}{}^{i_2...i_p}-\frac{\hat{g}_{ij}}{4p!}\hat{H}_{(m)}^2\right)\\
\hat{\epsilon}^1_{ij}&=-\frac{1}{2}\hat{g}_{ij}+2\alpha_2\hat{G}_{ij}^p+3\alpha_3\hat{\mathcal{H}}_{ij}^p+\beta_1\left(p\hat{H}_{(1)ii_2...i_p}\hat{H}_{(1)j}{}^{i_2...i_p}-\frac{\hat{g}_{ij}}{2}\hat{H}_{(1)}^2\right)\\
\hat{\epsilon}^2_{ij}&=-\frac{1}{2}\alpha_2\hat{g}_{ij}+3\alpha_3\hat{G}_{ij}^p+\beta_2\left(p\hat{H}_{(2)ii_2...i_p}\hat{H}_{(2)j}{}^{i_2...i_p}-\frac{\hat{g}_{ij}}{2}\hat{H}_{(2)}^2\right)\\
\hat{\epsilon}^{em}_{ij}&=2\beta_0p!\left(p\hat{H}_{(3)ii_2...i_p}\hat{H}_{(3)j}{}^{i_2...i_p}-\frac{\hat{g}_{ij}}{2}\hat{H}_{(3)}^2\right)-\frac{1}{8}\hat{g}_{ij}
\end{align}
on the internal manifold. 
Both sets of field equations impose compatibility conditions, that when written in terms of the magnetic charges and the bare cosmological constant, take the form
\begin{align}
&Q_{m_1}^2 = -\frac{6\alpha_3(p-1)(p-2)(p-3)(p-6+d)\gamma^2}{p!\beta_1(d-4)}-\frac{2\alpha_2(p-1)(2p-6+d)\gamma}{p!\beta_1(d-4)}-\frac{2}{p!\beta_1(d-4)}\nonumber \\
& Q_{m_2}^2 = -\frac{3\alpha_3(p-1)(p-6+d)\gamma}{p!\beta_2(d-5)}-\frac{\alpha_2}{p!\beta_2(d-5)}\nonumber \\
&Q_{m_3}^2 = -\frac{1}{8\beta_0p!^2(d-5)},  \label{CCC1}
\end{align}  
\begin{align}
&\Lambda = \frac{1}{2}\alpha_3(p-5)(p-1)(p-2)(p-3)(p-4)(d+p-6)\gamma^3\nonumber \\
&+\left(\frac{1}{6}(p-1)(p-2)(p-3)(3p-12+2d)\alpha_2+\frac{1}{2}\frac{(p-1)(p-2)(p-3)(d-3)(p-6+d)\alpha_3}{p!\beta_1(d-4)}\right)\gamma^2\nonumber \\
&+\left(\frac{1}{6}\frac{(p-1)(d-3)(2p-6+d)\alpha_2}{p!\beta_1(d-4)}+\frac{1}{4}\frac{(p-1)(d-3)(p-6+d)\alpha_3}{p!\beta_2(d-5)}+\frac{1}{6}(p-1)(3p-6+d)\right)\gamma\nonumber \\
&+\frac{1}{12}\frac{(d-3)\alpha_2}{p!(d-5)\beta_2}+\frac{1}{6}\frac{d-3}{p!(d-4)\beta_1}+\frac{1}{96}\frac{(d-3)}{(d-5)p!^2\beta_0}.  \label{CCC2}
\end{align}
Here we have explicitly introduced the quantities 
\begin{align}
\hat{R}_p&=\gamma p(p-1)\\
\hat{\mathcal{L}}^2_p&=\gamma^2p(p-1)(p-2)(p-3)\\
\hat{\mathcal{L}}^3_p&=\gamma^3p(p-1)(p-2)(p-3)(p-4)(p-5),
\end{align}
and the corresponding traces of Lovelock tensors of order $k$ 
\begin{equation}
g^{AB}E_{AB}^{k}=\left(\frac{2k-D}{2}\right)\mathcal{L}^k.
\end{equation}
Field equations (\ref{eqmc1}) and (\ref{eqmc2}) naturally contain the field equations of the 
Einstein-Maxwell-Gauss-Bonnet theory consider in Sec. IV-A for $\hat{H}_{(2)}=\hat{H}_{(3)}=0$  and $\alpha_3=0$. According to this, the compatibility conditions take the form
\small
\begin{align}
Q_{m_1}^2 &= -\frac{2\alpha_2}{\beta_1}\frac{(p-1)(d+p-4)}{p!(d-3)}\gamma-\frac{1}{p!\beta_1(d-3)}\label{cc1}\\
\Lambda &= \frac{1}{2}\alpha_2(p-1)(p-2)(p-3)(d+p-4)\gamma^2+\frac{1}{4}(p-1)(d+2p-4)\gamma-\frac{1}{8}Q_{m_1}^2(d-2)\label{cc2}.
\end{align}

\end{document}